\documentclass[12pt]{article}
\usepackage[cp1251]{inputenc}
\usepackage[russian]{babel}
\begin{document}
\large
\begin{center}
{\bf Schemes and Mechanisms of Neutrino Mixings (Oscillations) and
a Solution of the Sun Neutrinos Deficit Problem} \vspace{0.3cm}
\par
Beshtoev Kh. M.
\par
Joint Institute for Nuclear Research, Joliot Curie 6, 141980
Dubna, Moscow region, Russia \\
\end{center}

\vspace{0.5cm}
\par
{\bf Abstract}\\
\par
Three schemes of neutrino mixings (oscillations) are proposed. The
problems of origin of angle mixings, with the law of
energy-momentum conservation and disintegration of neutrino as
wave pocket are solved. These two schemes belong to mass mixings
schemes, where mixing angles and oscillation lengths are expressed
via elements of mass matrix. The third scheme belongs to the
charge mixings scheme, where mixing parameters are expressed via
neutrino weak charges, as it takes place in the vector dominance
model. Using experiments we must decide which of these schemes is
realized indeed. Analysis of the resonance enhancement mechanism
of neutrino oscillations in matter is performed. It is shown that
there are no indications on existence of this effect. It is shown
that the supposition that the neutrinos are Majorana particles is
not confirmed by accelerator experiments. Then only mixings
(oscillations) between Dirac neutrinos with different flavors
without sterile neutrinos can be realized. Using all the present
experimental data and the theoretical results the problem of Sun
neutrinos deficit is analyzed. The conclusion is: the primary Sun
$\nu_e$ neutrinos are converted into mixtures of three types of
neutrinos $\nu_e, \nu_\mu, \nu_\tau$ in approximately
equal quantities. \\

\par
{\bf I. Introduction}\\

\par
In the old theory of neutrino oscillations [1, 2], constructed in
the framework of Quantum Mechanics by analogy with the theory of
$K^{o}, \bar{K}^{o}$ oscillation, it is supposed that mass
eigenstates are $\nu_{1}, \nu_{2}, \nu_{3}$ neutrino states but
not physical neutrino states $\nu_{e}, \nu_{\mu }, \nu_{\tau}$,
and that the neutrinos $\nu_{e}, \nu_{\mu }, \nu_{\tau}$ are
created as superpositions of $\nu_{1}, \nu_{2}, \nu_{3}$  states.
This meant that the $\nu_{e}, \nu_{\mu }, \nu_{\tau}$ neutrinos
have no definite mass; i.e., their masses may vary depending on
the  $\nu_{1}, \nu_{2}, \nu_{3}$ admixture in the $\nu_{e},
\nu_{\mu }, \nu_{\tau}$  states. It is clear that this picture is
incorrect.
\par
Originally, it was supposed [2] that these neutrino oscillations
are real oscillations; i.e., that real transition of electron
neutrino $\nu_e$ into muon neutrino $\nu_{\mu}$ (or tau neutrino
$\nu_{\tau}$) takes place. Then the neutrino $x = \mu, \tau$
decays in electron neutrino plus something
$$
\nu_{x} \rightarrow \nu_e + ....  , \eqno(1)
$$
as a result, we get energy from vacuum, which equals the mass
difference (if $m_{\nu_x} > m_{\nu_e}$)
$$
\Delta E \sim m_{\nu_{x}} - m_{\nu_e} . \eqno(2)
$$
Then, again this electron neutrino transits into muon  neutrino,
which decays again and we get energy, and etc. {\bf So we got a
perpetuum mobile!} Obviously, the law of energy conservation
cannot be fulfilled in this process. The only way to restore the
law of energy conservation is to demand that this process is the
virtual one. Then, these oscillations will be the virtual ones and
they are described in the framework of the uncertainty relations.
The correct theory of neutrino oscillations can be constructed
only  into the framework of the particle physics theory, where the
conception of mass shell is present [3]-[6].
\par
i) If the masses of the $\nu_{e}, \nu_{\mu }, \nu_{\tau}$
neutrinos  are equal, then the real oscillation of the neutrinos
will take  place.
\par
ii) If  the masses  of  the $\nu_{e}, \nu _{\mu }, \nu _{\tau}$
are  not equal, then the virtual oscillation of  the  neutrinos
will  take place. To make these oscillations  real,  these
neutrinos must participate  in the quasi-elastic interactions, in
order to undergo transition  to the mass shell of other
appropriate neutrinos by analogy with $\gamma  - \rho ^{o}$
transition  in the  vector   meson  dominance model.
\par
At first we consider three schemes of neutrino mixings
(oscillations), then analyse main mechanisms of neutrino mixings
(oscillations) and then come to a solution of the Sun neutrinos
deficit problem using all available experimental data on
neutrino mixings. \\

\par
{\bf II. Schemes of Neutrino Mixings (Oscillations)} \\

\par
The mass matrix of $\nu_e$ and $\nu_\mu$ neutrinos has the
following form:
$$
\left(\begin{array}{cc} m_{\nu_e}& 0 \\ 0 & m_{\nu_\mu}
\end{array} \right) .
\eqno(3)
$$
\par
Due to the presence of the interaction violating the lepton
numbers, a nondiagonal term appears in this matrix and then this
mass matrix is transformed into the following nondiagonal matrix
($CP$ is conserved):
$$
\left(\begin{array}{cc}m_{\nu_e} & m_{\nu_e \nu_\mu} \\ m_{\nu_\mu
\nu_e} & m_{\nu_\mu} \end{array} \right) , \eqno(4)
$$
then the lagrangian of mass of the neutrinos takes the following
form ($\nu \equiv \nu_L$):
$$
\begin{array}{c}{\cal L}_{M} = - \frac{1}{2} \left[m_{\nu_e}
\bar \nu_e \nu_e + m_{\nu_\mu} \bar \nu_{\mu} \nu_{\mu } +
m_{\nu_e \nu_{\mu }}(\bar \nu_e \nu_{\mu } + \bar \nu_{\mu }
\nu _e) \right] \equiv \\
\equiv  - \frac{1}{2} (\bar \nu_e, \bar \nu_\mu)
\left(\begin{array}{cc} m_{\nu_e} & m_{\nu_e \nu_{\mu }} \\
m_{\nu_{\mu} \nu_e} & m_{\nu_\mu} \end{array} \right)
\left(\begin{array}{c} \nu_e \\ \nu_{\mu } \end{array} \right)
\end{array} ,
\eqno(5)
$$
which is diagonalized by turning through  the angle $\theta$ and
(see ref. in [2]) and then this lagrangian (5) transforms into the
following one:
$$
{\cal L}_{M} = - \frac{1}{2} \left[ m_{1} \bar \nu_{1} \nu_{1} +
m_{2} \bar \nu_{2} \nu_{2} \right]  , \eqno(6)
$$
where
$$
m_{1, 2} = {1\over 2} \left[ (m_{\nu_e} + m_{\nu_\mu}) \pm
\left((m_{\nu_e} - m_{\nu_\mu})^2 + 4 m^{2}_{\nu_\mu \nu_e}
\right)^{1/2} \right] ,
$$
\par
\noindent and angle $\theta $ is determined by the following
expression:
$$
tg 2 \theta  = \frac{2 m_{\nu_e \nu_\mu}} {(m_{\nu_\mu} -
m_{\nu_e})} , \eqno(7)
$$
$$
\begin{array}{c}
\nu_e = cos \theta  \nu_{1} + sin \theta \nu_{2}  ,         \\
\nu _{\mu } = - sin \theta  \nu_{1} + cos \theta  \nu_{2} .
\end{array}
\eqno(8)
$$
From Exp.(7) one can see that if $m_{\nu_e} = m_{\nu_{\mu}}$, then
the mixing angle is equal to $\pi /4$ independently of the value
of $m_{\nu_e \nu_\mu}$:
$$
sin^2 2\theta = \frac{(2m_{\nu_{e} \nu_{\mu}})^2} {(m_{\nu_e} -
m_{\nu_\mu})^2 +(2m_{\nu_e \nu_{\mu}})^2} , \eqno(9)
$$
$$
\left(\begin{array}{cc} m_{\nu_1} & 0 \\
0 & m_{\nu_2} \end{array} \right) .
$$

\par
It is interesting to remark that expression (9) can be obtained
from the Breit-Wigner distribution [7]
$$
P \sim \frac{(\Gamma/2)^2}{(E - E_0)^2 + (\Gamma/2)^2}   ,
\eqno(10)
$$
by using the following substitutions:
$$
E = m_{\nu_e},\hspace{0.2cm} E_0 = m_{\nu_\mu},\hspace{0.2cm}
\Gamma/2 = 2m_{\nu_e, \nu_\mu} ,
$$
where $\Gamma/2 \equiv W(... )$ is a width of $\nu_e \rightarrow
\nu_\mu$ transition, then we can use a standard method [4, 8] for
calculating this value.
\par
We can also see that there are two cases of $\nu_e, \nu_\mu$
transitions (oscillations) [4]-[6].
\par
1. If we consider the transition of $\nu_e$ into $\nu_\mu$
particle, then
$$
sin^2 2\theta \cong \frac{4m^2_{\nu_e, \nu_\mu}}{(m_{\nu_e} -
m_{\nu_\mu})^2 + 4m^2_{\nu_e, \nu_\mu}}  , \eqno(11)
$$
\par
How can we understand this  $\nu_e \rightarrow \nu_\mu$
transition?
\par
If $2m_{\nu_e, \nu_\mu} = \frac{\Gamma}{2}$ is not zero, then it
means that the mean mass of $\nu_e$ particle is $m_{\nu_e}$ and
this mass is distributed by $sin^2 2\theta$ (or by the
Breit-Wigner formula) and the probability of the $\nu_e
\rightarrow \nu_\mu$ transition differs from zero and it is
defined by masses of $\nu_e$ and $\nu_\mu$ particles and
$m_{\nu_e, \nu_\mu}$, which is computed in the framework of the
standard method, as pointed out above.
\par
Another interpretation of Exp. (11) is: If $m_{\nu_e, \nu_\mu}$
differs from zero then Exp. (11) gives probability of $\nu_e
\leftrightarrow \nu_\mu$ transitions. If $m_{\nu_e, \nu_\mu} = 0$,
then the ${\nu_e \leftrightarrow \nu_\mu}$ transitions are
forbidden.
\par
So, this is a solution of the problem of the of mixing angle
origin in the theory of vacuum oscillations.
\par
In this case the probability of $\nu_e \rightarrow \nu_\mu$
transition (oscillation) is described by the following expression:
$$
P(\nu_e \rightarrow \nu_\mu, t) =  sin^2 2\theta sin^2 \left[\pi
t\frac{\mid m_{\nu_1}^2 - m_{\nu_2}^2 \mid}{2 p_{\nu_e}} \right ],
\eqno(12)
$$
$$
L_{o} = 2\pi  {2p \over {\mid m^{2}_{2} - m^{2}_{1} \mid}} ,
$$
where $p_{\nu_e}$ is a momentum of $\nu_e$ neutrino and $L_{o}$ is
the length of neutrino oscillations.
\par
2. If we consider the virtual transition of $\nu_e$ into $\nu_\mu$
neutrino at  $m_{\nu_e} = m_{\nu_\mu}$ (i.e., without changing the
mass shell), then
$$
tg 2\theta = \infty  , \eqno(13)
$$
$\theta = \pi/4$, and
$$
sin^2 2\theta = 1     . \eqno(14)
$$
\par
In this case the probability of the $\nu_e \rightarrow \nu_\mu$
transition (oscillation) is described by the following expression:
$$
P(\nu_e \rightarrow \nu_\mu, t) = sin^2 \left[\pi t\frac{4
m_{\nu_e, \nu_\mu}^2}{2 p_a} \right ] . \eqno(15)
$$
\par
In order to make these virtual oscillations real, their
participation in quasi-elastic interactions is necessary for the
transitions to their own mass shells [8].
\par
It is clear that the $\nu_e \rightarrow \nu_\mu$ transition is a
dynamical process.
\par
3. The third type of transitions (oscillations) can be realized by
mixings of the fields (neutrinos) by analogy with the  vector
dominance model ($\gamma-\rho^o$ and $Z^o-\gamma$  mixings) in a
way as it takes place in the particle physics. Since the weak
couple constants $g_{\nu_e}, g_{\nu_{\mu}},  g_{\nu_{\tau}}$ of
$\nu_e, \nu_{\mu}, \nu_{\tau}$  neutrinos are nearly equal in
reality, i.e., $g_{\nu_e}\simeq g_{\nu_{\mu}} \simeq
g_{\nu_{\tau}}$  the angle mixings are nearly maximal:
$$
sin \theta_{\nu_{e} \nu_{\mu}} \simeq
\frac{g_{\nu_e}}{\sqrt{g^2_{\nu_{e}} + g^2_{\nu_{\mu}}}}
=\frac{1}{\sqrt{2}} \simeq sin \theta_{\nu_{e} \nu_{\tau}} \simeq
sin \theta_{\nu_{\mu} \nu_{\tau}}. \eqno(16)
$$
Therefore, if the masses of these neutrinos are  equal (which is
hardly probable), then  transitions between neutrinos will be real
and if the masses of these  neutrinos are not equal, then
transitions between  neutrinos will be virtual by analogy with
$\gamma-\rho^o$ transitions. In this approach we can hardly see
neutrino oscillations, though it is possible, in principle.
\par
Probably, a more realistic type of neutrino transitions is the
third one since oscillations can arise by a dynamics. Exactly the
third type of mixings is realized through dynamical charges
$g_{\nu_e}, g_{\nu_{\mu}}, g_{\nu_{\tau}}$ of neutrinos  but not
through neutrino masses. Nevertheless, we must get an answer to
the question: which of the above three types of neutrino
oscillations
is realized in the Nature?\\

{\bf III. Mechanisms of Neutrino Oscillations}\\

\par
{\bf III.1. Impossibility of Resonance Enhancement of Neutrino
\par
Oscillations in Matter}\\

\par
Via three different approaches: by using mass Lagrangian [5],
[9,10], by using the Dirac equation [5, 11], and by using the
operator formalism [12], the author of this work has discussed the
problem of the mass generation in the standard weak interactions
and has come to a conclusion that the standard weak interaction
cannot generate masses of fermions since the right-handed
components of fermions do not participate in these interactions.
It is also shown [13] that the equation for Green function of the
weak-interacting fermions (neutrinos) in the matter coincides with
the equation for Green function of fermions in vacuum and the law
of conservation of the energy and the momentum of neutrino in
matter will be fulfilled [12] only if the energy $W$ of
polarization of matter by the neutrino or the corresponding term
in Wolfenstein equation [14], is zero (it means that neutrinos
cannot generate permanent polarization of matter). These results
lead to the conclusion:  resonance enhancement of neutrino
oscillations in matter does not exist.
\par
The simplest method to prove the absence of the resonance
enhancement of neutrino oscillations in matter is:
\par
   If we put an electrical ($e$) (or strong ($g$)) charged particle $a$ in  vacuum,
there arises polarization of vacuum. Since the field around
particle $a$ is spherically symmetrical, the polarization must
also be spherically symmetrical. Then the particle will be left at
rest and the law of energy and momentum conservation is fulfilled.
\par
If we put a weakly ($g_W$) interacting particle $b$ (a neutrino)
in vacuum, then since the field around the particle has a
left-right asymmetry (weak interactions are left-handed
interactions with respect to the spin direction [15, 16]),
polarization of vacuum must be nonsymmetrical; i.e., on the left
side there arises maximal polarization and on the right there is
zero polarization. Since polarization of the vacuum is
asymmetrical, there arises asymmetrical interaction of the
particle (the neutrino) with vacuum and the particle cannot be at
rest and will be accelerated. Then the law of energy momentum
conservation will be violated. The only way to fulfil the law of
energy and momentum conservation is to demand that polarization of
vacuum be absent in the weak interactions. The same situation will
take place in matter. It is necessary to remark that for the
above-considered proof it is sufficient to know that the field
around the weakly interacting particle is asymmetrical (and it is
not necessary to know the precise form of this field). It is
necessary also to remark that the Super-Kamiokande data on
day-night asymmetry [17] is
$$
A = (D - N)/(\frac{1}{2}(D + N)) = -0.021 \pm 0.020(stat) +
0.013(-0.012)(syst) . \eqno(17)
$$
and it does not leave hope on possibility of the resonance
enhancement of neutrino oscillations in matter.
\par
In means that the forward scattering amplitude of the weak
interactions has specific behavior.
\par
In the Sun neutrino experiments, it is impossible to distinguish
muon and tau neutrinos since energies of the Sun neutrinos are
less than the threshold energy of muon and tau lepton productions.
However, we can distinguish these neutrinos on the Earth-long-
baseline accelerator neutrino experiments with neutrino energies
higher than the threshold energy of muon and tau lepton creations.
Probably, such experiments put an end to the problem of resonance
enhancement of neutrino oscillations in matter (see the strong
proof absence of this effect given above). \\

\par
{\bf III.2. Majorana Neutrino Oscillations}\\
\par
At present it is supposed [18] that the neutrino oscillations can
be connected with Majorana neutrino oscillations. It will be shown
that we cannot put Majorana neutrinos in the standard Dirac
theory. It means that in experiments the Majorana neutrino
oscillations cannot be observed.
\par
Majorana fermion in Dirac representation has the following form
[1, 2, 19]:
\par
$$
\chi^M = \frac{1}{2} [\Psi(x) + \eta_C\Psi^{C}(x)] , \eqno(18)
$$
$$
\Psi^C(x) \rightarrow \eta_C C \bar \Psi^T(x) ,
$$
\par
\noindent where $\eta_{C}$ is a phase, $C$  is a charge
conjunction, $T$ is a transposition.
\par
In the standard theory all fermion are Dirac particles and this
theory is gauge invariance one. Then the supposition that neutrino
is a Majorana particle is a right violation of gauge invariance
[6, 20].
\par
If, in spite of the above arguments, we put Majorana neutrinos in
the standard theory, there appears two possibilities:
\par
1. If the second component of Majorana neutrino in Eq.(18) is
antineutrino, then the neutrinoless double beta decay will take
place, neutrino will transit into antineutrino while oscillations
and then in accelerator experiments we must see the following
reactions:
$$
\chi_l + A(Z) \to l^{-} + A(Z+1), \eqno(19)
$$
with relative probability 1/2 and
$$
\chi_l + A(Z) \to l^{+} + A(Z-1), \eqno(20)
$$
with the same relative probability (where $l = e, \mu, \tau$),
since Majorana neutrinos are superpositions of Dirac neutrinos and
antineutrinos.
\par
2. If the second component of Majorana neutrino in Eq.(18) is a
sterile neutrino, then the neutrinoless double beta decay cannot
exist, since the second component has incorrect spirality but we
can see neutrino disappears at oscillations and then in
accelerator experiments we must see the following reactions:
$$
\chi_l + A(Z) \to l^{-} + A(Z+1), \quad l = e, \mu, \tau ,
\eqno(21)
$$
with relative probability 1/2. The second sterile components of
Majorana neutrinos in Eq.(18) do not take part in the standard
weak interactions. Obviously, all the available accelerator
experimental data [21] do not confirm these predictions; therefore
we cannot consider this mechanism as a realistic one for neutrino
oscillations. Then, in principle, transition of the Dirac
neutrinos into of the Majorana neutrinos can be, but it is only
possible at full violation of the lepton numbers, i.e., at the
Grand Unification scales ($T > 10^{30}$ y.). \\

\par
{\bf III.3. Mixings (Oscillations) of Flavor Neutrinos}\\
\par
 In the work [22] Z. Maki et al. and B. Pontecorvo in [23]
 supposed that there  could exist
transitions between aromatic neutrinos $\nu_e, \nu_\mu$.
Afterwards $\nu_\tau$ was found and then $\nu_e, \nu_\mu,
\nu_\tau$ transitions could be possible. It is necessary to remark
that only this scheme of oscillations is realistic for neutrino
oscillations. The expressions, which described neutrino
oscillations in this case are given above
in expressions (11)-(17).\\

\par
{\bf IV. A Solution of the Sun Neutrinos Deficit Problem} \\

\par
Before consideration of this problem it is important to consider
the necessary data.
\par
The value of the Sun neutrinos flow measured (through elastic
scattering) on SNO [24] is in good agreement with the same value
measured in Super-Kamiokande [25].
\par
Ratio of $\nu_e$ flow measured on SNO (CC) to the same flow
computed in the frame work of SSM [26] ($E_\nu > 6.0 MeV$) is:
$$
\frac{\phi_{SNO}^{CC}}{\phi_{SSM2000}} = 0.35 \pm 0.02 . \eqno(22)
$$
This value is in good agreement with the same value of $\nu_e$
relative neutrinos flow measured on Homestake (CC) [27] for energy
threshold $E_\nu = 0,814 MeV$.
$$
 \frac{\Phi^{exp}}{\Phi^{SSM2000}} = 0.34 \pm 0.03 .
\eqno(23)
$$
From these data we can come to a conclusion that the angle mixing
for the Sun $\nu_e$ neutrinos does not depend on neutrino energy
thresholds. Now it is necessary to know the value of this angle
mixing $\theta_{\nu_e \nu_\mu}$. Estimation of the value of this
angle can be extracted from KamLAND [28] data and it is:
$$
sin^2 \theta_{\nu_e \nu_\mu} \cong 1.0 , \quad \theta \cong
\frac{\pi}{4}, \eqno(24)
$$
The angle mixing for vacuum $\nu_\mu \to \nu_\tau$ transitions
obtained on Super-Kamiokande [29] for atmospheric neutrinos is:
$$
sin^2 2\theta_{\nu_\mu \nu_\tau}  \cong  1, \quad \theta \cong
\frac{\pi}{4} . \eqno(25)
$$
Now we can estimate the third angle mixing for ${\nu_e \to
\nu_\tau}$ transitions. The full flow neutrinos obtained on SNO
[24] is:
$$
\phi_{SNO} (\nu_x) = (5.09 \pm 0.44) \times 10^6 cm^{-2}c^{-1} .
\eqno(26)
$$
This result is in good agreement with the following prediction of
the Standard Sun Model [26]:
$$
\phi_{SNO} (\nu_x) = (5.05 + 1.01( -0.81)) \times 10^6
cm^{-2}c^{-1} . \eqno(27)
$$
This agreement is an indication on  absence of neutrino
disappearance, i. e. the sterile neutrino hypothesis is not
confirmed.
\par
The SNO experiment has measured only the summary flow of $\mu_\mu
+ \nu_\tau$ neutrinos. Now we must separate these flows. For this
purpose we can use neutrino mixing parameters obtained on
Super-Kamiokande for atmospheric neutrinos (25), on KamLAND (24)
for reactor antineutrinos and also the following flow of the Sun
electron neutrinos measured on SNO [24]:
$$
\phi^{CC}_{SNO} (\nu_e) = (1.76\pm 0.11) \cdot 10^6 cm^{-2} s^{-1}
. \eqno(28)
$$
In the subsequent considerations we will take into account that in
the Sun neutrino experiments where the measured neutrinos with
energy up to 15 MeV, and that the Earth orbit is not a circular
one i.e., the measured values for the Sun neutrinos are average
values.
\par
So, since the obtained vacuum angle mixing for $\nu_e \to \nu_\mu$
is close to the maximal angle then the sum $\nu_e + \nu_\mu$ Sun
neutrino flow is a doubled flow of $\nu_e$ Sun neutrinos
$$
\phi_{SNO} (\nu_e + \nu_\mu) \cong (3.52\pm 0.22) \cdot 10^6
cm^{-2} s^{-1} . \eqno(29)
$$
Then the following remainder of the Sun neutrinos flow is the flow
of $\nu_\tau$ Sun neutrinos:
$$
\phi_{SNO} (\nu_\tau) \cong \phi_{SNO} (\nu_x) - \phi_{SNO} (\nu_e
+ \nu_\mu) \cong  (1.57 \pm 0.49) \cdot 10^6 cm^{-2}c^{-1} .
\eqno(30)
$$
It is about one third of the primary Sun neutrino flow. Since the
angle mixings of $\theta_{\nu_e \nu_\mu}$ and  $\theta_{\nu_\mu
\nu_\tau}$ neutrinos are close to maximal angles, then we have no
reason to suppose that the $\theta_{\nu_e \nu_\mu}$ angle mixing
distinctly differs from the maximal one. So we come to the
following conclusion: on way to the Earth the primary Sun $\nu_e$
neutrinos transit in the mixture of electron, muon and tau
neutrinos in approximately equal quantities:
$$
\phi_{Sun} (\nu_e) \to \phi_{Sun} (\nu) \cong \frac{1}{3}
\phi(\nu_e) + \frac{1}{3} \phi(\nu_\mu) + \frac{1}{3}
\phi(\nu_\tau) . \eqno(31)
$$
This is a decision of the Sun neutrinos deficit problem at the
qualitative level i.e., the vacuum mixing angles are near to the
maximal ones and there are no serious indication on existence of
sterile neutrinos. Besides there are no experimental indication on
realization of the mechanism of resonance enhancement of neutrino
oscillations in matter.  It is necessary, in subsequent long
baseline experiments on the Earth, to precise all neutrino angle
mixings and squared mass differences of neutrinos for their using
for precise solution of the Sun neutrinos deficit problem. The
problem about existence or absence of neutrino oscillations is
very important. \\

{\bf IV. Conclusion} \\

\par
Three schemes of neutrino mixings (oscillations) has been
proposed. The first scheme is a development of the standard
schemes in the framework of particle physics, where problems of
origin of angle mixings, with the law of energy-momentum
conservation and disintegration of neutrino as wave pocket are
solved. In the second scheme the angle mixing is maximal. These
two schemes belong to mass mixings schemes, where mixing angles
and oscillation lengths are expressed via elements of mass matrix.
The third scheme belongs to the charge mixings scheme, where
mixing parameters are expressed via neutrino weak charges, as it
takes place in the vector dominance model. Using experiments we
must decide which of these schemes is realized indeed. Analysis of
the resonance enhancement mechanism of neutrino oscillations in
matter is performed. It is shown that there are no indications on
existence of this effect. It is shown that the supposition that
the neutrinos are Majorana particles has not been confirmed by
accelerator experiments. Then only mixings (oscillations) between
Dirac neutrinos with different flavors without sterile neutrinos
can be realized. Using all the present experimental data and the
theoretical results the problem of Sun neutrinos deficit has been
analyzed. The conclusion is: the primary Sun $\nu_e$ neutrinos are
converted into mixtures of three types of neutrinos $\nu_e,
\nu_\mu, \nu_\tau$ in approximately equal quantities. \\

\par
{\bf References}\\

\par
\noindent 1. Gribov V., Pontecorvo B. M., Phys. Lett. B, 1969,
v.28, p.493.
\par
\noindent 2. Bilenky S. M., Pontecorvo B. M., Phys. Rep. C, 1978,
v.41, p.225;
\par
Boehm F., Vogel P., Physics of Massive Neutrinos: Cambridge
\par
Univ. Press, 1987, p.27, p.121;
\par
Bilenky S. M., Petcov S. T., Rev. of Mod.  Phys., 1977, v.59,
\par
p.631.

\par
\noindent 3. Beshtoev Kh. M., JINR Commun. E2-92-318, Dubna, 1992;
\par
JINR Rapid Communications, N3[71]-95.
\par
\noindent 4. Beshtoev Kh. M., hep-ph/9911513;
\par
 The Hadronic Journal, 2000, v.23, p.477;
\par
Proceedings of 27th Intern. Cosmic Ray Conf., Germany,
\par
Hamburg, 7-15 August 2001, v.3, p. 1186.
\par
\noindent 5. Beshtoev Kh. M., Phys. of Elem. Part. and Atomic
Nucl.
\par
(Particles and Nuclei), 1996, v.27, p.53.
\par
\noindent 6. Beshtoev Kh. M., Proceed. of 28th International
Cosmic Ray Conference,
\par
Japan, 2003, V.3, p.1503; Japan, 2003, V.3, p.1507;   Japan, 2003,
V.3,
\par
p.1511; JINR Commun. E2-2004-58, Dubna, 2004.
\par
\noindent 7. Blatt J. M., Waiscopff V. F., The Theory of Nuclear
Reactions,
\par
INR T.R. 42.
\par
\noindent 8. Beshtoev Kh. M., JINR Commun. E2-99-307, Dubna, 1999;
\par
JINR Commun. E2-99-306, Dubna, 1999.
\par
\noindent 9. Beshtoev Kh. M., JINR Commun. E2-91-183, Dubna, 1991;
\par
Proceedings of III  Int. Symp. on Weak and Electromag. Int. in
\par
Nucl. (World Scient., Singapoure, p.781, 1992);
\par
\noindent 10. Beshtoev Kh. M., JINR Communication E2-93-167,
Dubna,
\par
1993; JINR Communication P2-93-44, Dubna, 1993;
\par
\noindent 11. Beshtoev Kh. M., JINR Communication E2-93-167,
Dubna,
\par
1993; JINR Communication P2-93-44, Dubna, 1993.
\par
\noindent 12. Beshtoev Kh. M., hep-ph/9912532, 1999;
\par
Hadronic Journal, 1999, v.22, p.235.
\par
\noindent 13. Beshtoev Kh. M., JINR Communication E2-2000-30,
Dubna,
\par
2000; hep-ph/0003274.
\par
\noindent 14. Wolfenstein L., Phys. Rev. D, 1978, v.17, p.2369;
\par
 Mikheyev S. P., Smirnov A. Ju., Nuovo Cimento, 1986, v.9, p.17.
\par
\noindent 15. Glashow S. L.,  Nucl. Phys., 1961, v.22, p.579 ;
\par
Weinberg S., Phys.  Rev. Lett., 1967, v.19, p.1264 ;
\par
Salam A.,  Proc. of the 8th Nobel  Symp.,  edited  by
\par
N. Svarthholm (Almgvist and Wiksell,  Stockholm) 1968, p.367.
\par
\noindent 16. Beshtoev Kh. M., JINR Communication D2-2001-292,
Dubna,
\par
2001; hep-ph/0103274.
\par
\noindent 17. Kameda J., Proceedings of ICRC 2001, August 2001,
Germany,
\par
Hamburg, p.1057.
\par
Fukuda  S. et al,. Phys.   Rev. Lett., 2001, v.25, p.5651;
\par
Phys. Lett. B, 539, 2002,  p.179.
\par
\noindent 18. Gonzalez-Garcia C., 31-st ICHEP, Amsterdam, July
2002.
\par
\noindent 19. Rosen S. P., Lectore Notes on Mass Matrices, LASL
preprint,
\par
1983.
\par
\noindent 20. Beshtoev Kh. M., JINR Communic. E2-2003-155, Dubna,
2003; hep-ph/0212210, 2002;  hep-ph/0304157.
\par
\noindent 21. Phys. Rev D66, N1, 2002, Review of Particle Physics.
\par
\noindent 22. Maki Z. et al., Prog. Theor. Phys., 1962, v.28,
p.870.
\par
\noindent 23. Pontecorvo B. M., Soviet Journ. JETP, 1967, v.53,
p.1717.
\par
\noindent 24. Ahmad Q. R. et al., Internet Pub. nucl-ex/0106015,
June 2001.
\par
Ahmad  Q. R. et al., Phys. Rev. Lett. 2002, v. 89, p.011301-1;
\par
Phys. Rev. Lett.  2002,v.  89, p.011302-1.
\par
\noindent 25. Fukuda S. et al., Phys Rev. Lett., 2001, v.25,
p.5651;
\par
Phys. Lett.B, 2002, v.539 p.179.
\par
Koshio Y. (Super-Kamiokande Collab.), Proc. of 28-th Intern.
Cosmic Ray Conf., Japan, 2003, v.1, p.1225.
\par
\noindent 26. Bahcall D. et al., The Astrophysical Jour. 2001,
v.555, p.990.
\par
\noindent 27. Davis R., Prog. Part. Nucl. Phys., 1994, v.32, p.13
\par
\noindent 28. Eguchi K. et al., Phys. Rev. Let. 2003, v.90,
021802;
\par
Mitsui T., 28-th Intern. Cosmic Ray Conf., Japan, 2003, v.1
p.1221.
\par
\noindent 29. Fukuda S. et al., Phys Rev. Lett. 2001, v.25,
p.5651;
\par
Habig A., Proceedings of Inter. Cosmic Ray Conf., Japan, 2003,
v.1, p. 1255.

\end{document}